\documentclass{aastex}
\usepackage{spr-astr-addons}
\usepackage{url}

\RequirePackage{color}

\begin{document}

\title{General Relativistic Magnetohydrodynamic and Monte Carlo Modeling of Sagittarius A*}
\slugcomment{Not to appear in Nonlearned J., 45.}
\shorttitle{GRMHD and MC Modeling of Sagittarius A*}
\shortauthors{Hilburn et al.}

\author{Guy Hilburn\altaffilmark{1,2}} 
\and 
\author{Edison Liang\altaffilmark{1}}
\and
\author{Siming Liu\altaffilmark{3}}
\and
\author{Hui Li\altaffilmark{2}}

\altaffiltext{1}{Department of Physics and Astronomy, Rice University, 6100 Main, Houston, Texas, 77005}
\altaffiltext{2}{Los Alamos National Laboratory, P.O. Box 1663, Los Alamos, New Mexico, 87545}
\altaffiltext{3}{Department of Physics and Astronomy, University of Glasgow, G12 8QQ, Scotland}

\begin{abstract}
We present results of models of the physical space and parameters of the accretion disk of Sagittarius A*, as well as simulations of its emergent spectrum.  This begins with HARM, a 2D general relativistic magneto-hydrodynamic (GRMHD) model, specifically set up to evolve the space around a black hole.  Data from HARM are then fed into a 2D Monte-Carlo (MC) code which generates and tracks emitted photons, allowing for absorption and scattering before they escape the volume.
\end{abstract}

\keywords{accretion disc; black hole; Galactic Center; magnetohydrodynamics}

\section{Introduction}

Sagittarius A* is the supermassive black hole (about 3.6 million Solar masses) at the center of our galaxy, which is known as a strong radio source with specific x-ray signatures which have come to be associated with flaring or quiescent states.  The short time scales of these flares, as well as the lack of significant time lags between NIR and x-ray components, suggests the possibility of these arising in a small volume close to the event horizon.

Recent attempts to model the spectrum of Sagittarius A* have proven promising, but have relied upon simplifications or assumptions which have somewhat diminished the impact of these results.  Our intent is to deliver the most consistent modeling possible with current technology, and avoid some of the simplifications that previous simulations have required.

This begins with using a GRMHD code specifically designed to evolve an accretion disk around a black hole.  This is a considerable advantage over models which specify analytic formulae to describe density, temperature, and magnetic field profiles.  Even simulations done with MHD codes suffer some of the same downfalls as simpler set-ups --- specifically that effects of general relativity near a supermassive black hole become quite important.

Data output by the GRMHD code are used as input for our MC code.  Unlike other models which rely only on emission and scattering calculated analytically, this simulation creates a realistic situation through use of the Monte-Carlo random generation method for photons, and allows for scattering and absorption before photons escape the volume.

It should be noted that both of the codes used here are 2D and axially symmetric.  The third dimension would be useful for full consistency, but the authors feel that its lack of inclusion does not preclude the results from being significant.

\section{HARM GRMHD Code}

Our GRMHD modeling is done using the HARM (High Accuracy Relativistic Magnetohydrodynamics) code developed by \cite{gam03}.  The code simulates GRMHD evolution through time by the use of particle number conservation, the four energy-momentum equations, the MHD stress-energy tensor, and the induction equation.  These equations take conserved hyberbolic forms, for easy integration:
\begin{equation}
  \frac{dU}{dt} + \frac{dF(U)}{dx^i} = S(U)
\end{equation}
where U, F, and S represent the set of conserved variables, fluxes, and sources, respectively.

HARM is set to first seed a torus around a black hole with an initial density and poloidal magnetic field, then evolve the space through time, tracking density, internal energy, 4-velocity, 4-magnetic field.  This is allowed to continue until the space reaches a ``steady state'' - of special note in the code output is the highly turbulent magnetic field, an artifact of the magneto-rotational instability, which is an important mechanism governing the transport of angular momentum in the disk.  This instability causes the turbulence which is likely to be at least partly responsible for the energizing of electrons in the disk, though in our MC trials, we only consider fully thermal distributions.

\section{MC Emission and Scattering Code}

The MC code used for photon emission and scattering has a long history and is discussed in a number of resources, such as \cite{can87,lia88,bot98}.  For a complete treatment, readers should see these papers.

In general, this code is a coupled MC/FP (Fokker-Planck) code.  For our intents, the FP evolution of the electron distribution was unnecessary at this stage, so it was turned off to allow a fixed temperature given by the HARM output.  The code is set up on a 2D axially-symmetric cylindrical grid, creating a (hollow or solid, depending on whether the inner radius is set to zero) cylindrical shape.  Each zone is assigned a density, ion and electron temperatures, magnetic field amplitude, and thermal and nonthermal distribution components.  For our purposes, this is simply set to be a Maxwellian, but the code allows power law nonthermal distributions as well.  The code allows emission from the volume and boundaries, and emitted photons are tracked and allowed to scatter or absorb. 

Output from the HARM code is used to assign the zone quantities.  That is, we specify a maximum density and the rest of the output scales to give us the ion temperature and saturated MRI magnetic field.  Assuming that inclusion of an initial toroidal magnetic fields would add to the final field, as this dimension is mostly unaffected by evolution, allows us to set the magnetic field, as long as it doesn't drop below the saturation value.  Similarly, as we have a specified ion temperature but no way to directly evaluate the electrons' acceleration, we set a global ratio between the two values.  That is, in one trial the electron temperature in a zone could be set to always be twice the value of the ion temperature in  the same zone.

\subsection{Flaring Results}

For all of the fits presented in figures below, the open circles are data points, triangles are upper limits, and the bowties denote the Chandra-obtained flaring and quiescent x-ray data points and slopes.

The first three figures show fits to the flaring x-ray point.  As shown, the flaring bowtie has almost zero slope associated with it.  In order to match this, there were several options.  Obviously, the bremsstrahlung component is very flat out to the high energy cut-off point.  It is also possible for Compton-scattered components to reach these energies and have flat spectra - trials are shown for fits using first the first-scattered bump, and secondly, the bump arising from photons that scatter twice off of hot electrons.

These three spectral components provide nearly equally-adequate fits, if only considering the flaring behavior, though the second Compton trial in Figure 3 seems to best fit the lower energy components.

It should be noted that effort was not focused on fitting the very low energy radio points.  It is expected that these arise primarily from a very large volume of low density, temperature, and magnetic field around the accretion disc, which is beyond the scale of this experiment, as our volume cuts off at r = 40M.

\begin{figure}
  \includegraphics[width=\columnwidth]{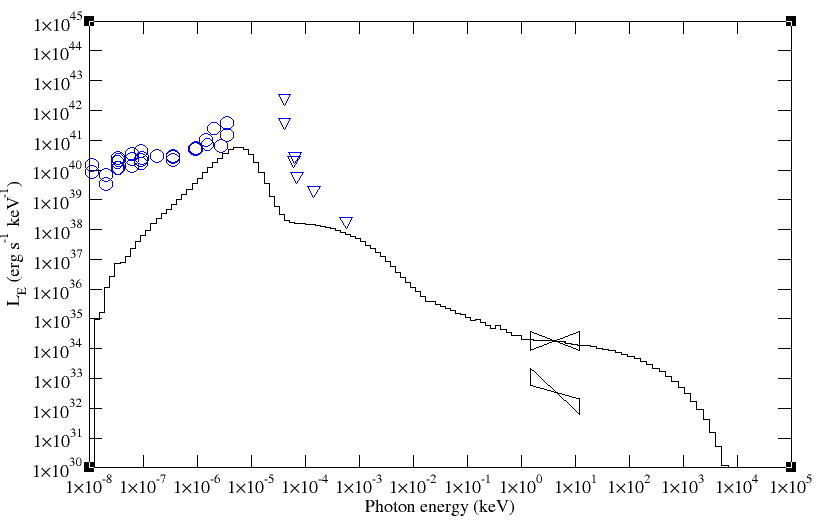}
  \caption{Fit to flaring data with bremsstrahlung component.  Maximum values of the 95x95 cell grid are scaled to n = 4.3x10$^{9}$ particles/cm$^{3}$, T = 1.2x10$^{3}$ keV, B = 9.9x10$^{2}$ G.  Total luminosity is 3.4x10$^{36}$ erg/s.}
\end{figure}

\begin{figure}
  \includegraphics[width=\columnwidth]{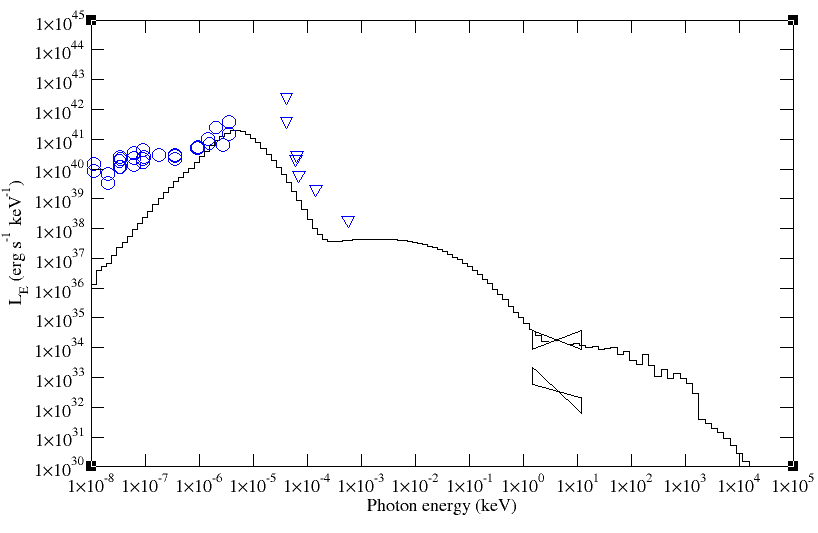}
  \caption{Fit to flaring data with second Compton bump component.  Maximum values of the 95x95 cell grid are scaled to n = 3.4x10$^{8}$ particles/cm$^{3}$, T = 8.2x10$^{3}$ keV, B = 1.6x10$^{2}$ G.  Total luminosity is 7.5x10$^{36}$ erg/s.}
\end{figure}

\begin{figure}
  \includegraphics[width=\columnwidth]{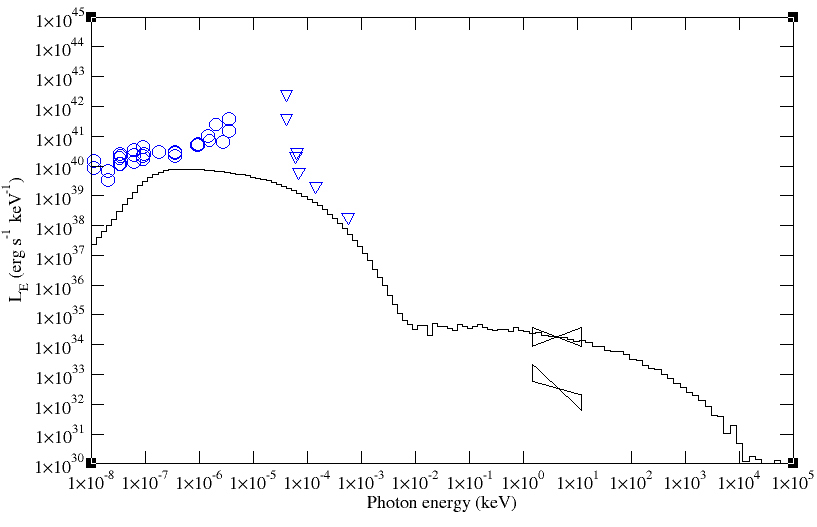}
  \caption{Fit to flaring data with first Compton bump component.  Maximum values of the 95x95 cell grid are scaled to n = 5.2x10$^{6}$ particles/cm$^{3}$, T = 1.8x10$^{5}$ keV, B = 13 G.  Total luminosity is 2.6x10$^{36}$ erg/s.}
\end{figure}

\subsection{Quiescent Results}

Once the flaring trials were completed, consideration had to be given to fitting the quiescent points with a consistent method, given the flaring fits.  It was found that very good fits to the quiescent point were possible by simply dropping the density or temperature independently, for the two Compton bump trials.  Unfortunately, the bremsstrahlung trial did not fit the quiescent point regardless of how its parameters were modified, as it will always have a very flat spectrum at the x-ray point.

The second Compton bump quiescent fit was achieved by changing only the density in the trial.  This is representative of a global mass accretion rate change, which could certainly be possible for observed variability from Sgr A*.

Conversely, it was found that for the first Compton bump trial, dropping the temperature allowed for a good fit.  This could indicate that the flaring state arises due to some large-scale magnetic energy dissipation event, such as rapid reconnection over a large area.

\begin{figure}
  \includegraphics[width=\columnwidth]{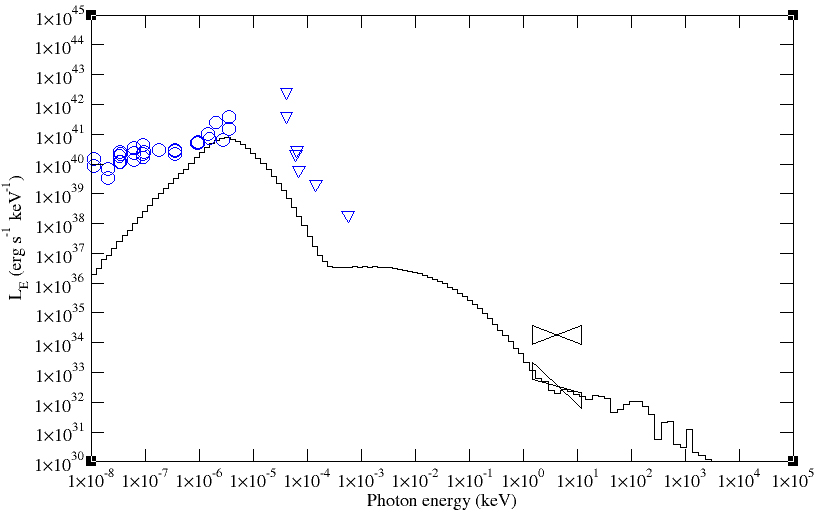}
  \caption{Fit to quiescent data with second Compton bump component.  Maximum values of the 95x95 cell grid are scaled to n = 6.9x10$^{7}$ particles/cm$^{3}$, T = 8.2x10$^{3}$ keV, B = 1.6x10$^{2}$ G.  Total luminosity is 8.2x10$^{35}$ erg/s.}
\end{figure}

\begin{figure}
  \includegraphics[width=\columnwidth]{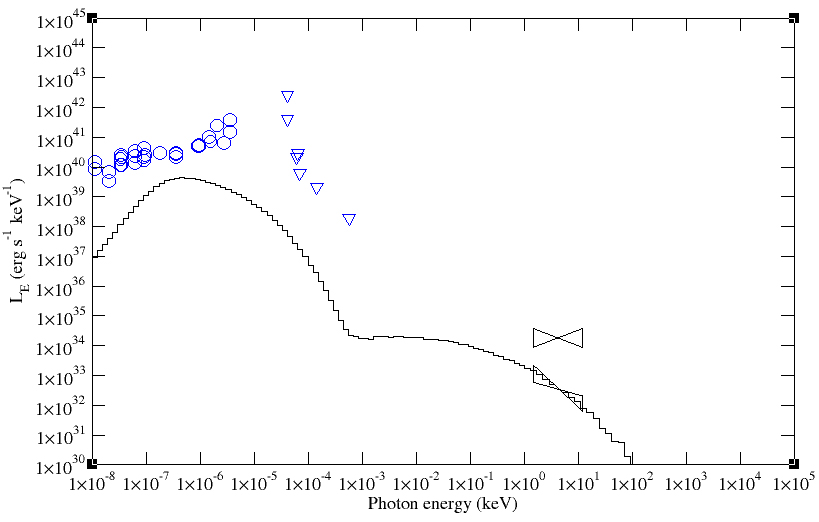}
  \caption{Fit to quiescent data with first Compton bump component.  Maximum values of the 95x95 cell grid are scaled to n = 5.2x10$^{6}$ particles/cm$^{3}$, T = 4.7x10$^{4}$ keV, B = 13 G.  Total luminosity is 3.4x10$^{34}$ erg/s.}
\end{figure}

\subsection{MC Code Modifications}

Since the initial project on fitting data from Sagittarius A* was completed, work has begun on modifying the Monte Carlo code to allow it to better undertake a number of problems.

First in this endeavor was rewriting the synchrotron emission routines of the code to support the inclusion of anisotropic magnetic fields.  As one may expect, in supermassive black hole accretion discs, the relative strengths of the components of the field differ greatly - being heavily biased toward the toroidal component.  Therefore, it is inconsistent to consider the emission from synchrotron radiation to be isotropic.  It should, instead, be heavily angle dependent.  Tests with this addition showed it to work as intended, and future projects will all include this feature.

As these simulations deal with situations in which individual regions of the volume may be moving with significantly different bulk velocities, work has begun on the inclusion into the MC code of differentially moving volume elements.  This will allow for another degree of consistency in results, by appropriately considering the boosting and Doppler shifting of photons emitted in each region.

\section{Conclusions}

Fits to the broadband spectra of Sgr A* have been shown using a combination of GRMHD accretion disc evolution and MC radiation transport codes.  Two sets of these results - those for the first and second Compton bump fits - show good fits to the quiescent x-ray point, and both sets, with reasonable changes in the physical parameters, fit the flaring x-ray point.  These changes are indicative of the near-global events, such as mass accretion increases, or the rapid dissipation of magnetic energy, which would be necessary to explain the observed variability.

The second Compton bump trials are presented as the most promising and complete detailed here.  The fact that these trials can closely match the IR and optical data, while fitting the x-ray points, makes these the most consistent with observations.

\acknowledgments
GH was supported by a LANL IGPP research grant and NSF AST-0406882.  EL was partially supported by NSF AST-0909167 and DOE- SC-0001481.

The authors would like to thank a number of people whose input helped shape this work, including, but not limited to, Markus B\"{o}ttcher, Xuhui Chen, Justin Finke, Charles Gammie, and Feng Yuan.

\bibliographystyle{spr-mp-nameyear-cnd}

\end{document}